\documentclass{article_saj}
\usepackage{cuted} 
\usepackage{graphicx,saj,multicol,subeqnarray}
\usepackage{natbib}
\usepackage{float}
\usepackage{xcolor}
\usepackage{url}
\usepackage{widetext}
\usepackage{bm}
\usepackage{tikz} % for checkmark
\usepackage{pifont} % for xmark
\usepackage{amsfonts}
\usepackage{amssymb}
\usepackage{amsmath,upgreek}
\usepackage{titlesec}
\usepackage[nolist]{acronym}
\usepackage{comment}

\def\point#1{\hbox{\setbox7=\hbox to0.6em{\hfil.\hfil}%
\setbox8=\hbox to0.5em{\hfil$^{#1}$\hfil}%
\box7\kern-0.5em\box8}}

\def\pointmin#1{\hbox{\setbox2=\hbox to0.8em{\hfil.\hfil}%
\setbox3=\hbox to0.6em{\hfil$^{#1}$\hfil}%
\box2\kern-.7em\box3}}

\def\mmm{\pointmin{\mathrm{m}}\kern.15em}

\titlelabel{\thetitle.\quad}
\definecolor{xlinkcolor}{cmyk}{1,0.6,0,0}
\usepackage[bookmarks=false,         % show bookmarks bar?
     pdfnewwindow=true,      % links in new window
     colorlinks=true,    % false: boxed links; true: colored links
     linkcolor=xlinkcolor,     % color of internal links
     citecolor=xlinkcolor,     % color of links to bibliography
     filecolor=xlinkcolor,  % color of file links
     urlcolor=xlinkcolor,      % color of external links
final=true
]{hyperref}

\def\papertype{\ \hfill\ Editorial}

\def\udc{52}
\newcommand{\ergcm}[1]{erg~cm$^{-2}$ s$^{-1}$}

\newcommand{\HI}{H{\sc i}}

\newcommand{\Halpha}{H${\alpha}$}

\newcommand{\D}{$^\circ$}

\newcommand{\cohen}{Cohen's~Ring}
\newcommand{\cohenstar}{Cohen's~Star}

\newcommand{\Ha}{H$\alpha$}

\def\arcmin{\hbox{$^\prime$}}
\def\arcsec{\hbox{$^{\prime\prime}$}}

\newcommand{\farcm}{\mbox{\ensuremath{.\mkern-4mu^\prime}}}%    % fractional arcminute symbol: 0.'0
\newcommand{\farcs}{\mbox{\ensuremath{.\!\!^{\prime\prime}}}}%  % fractional arcsecond symbol: 0.''0
\newcommand{\fdg}{\mbox{\ensuremath{.\!\!^\circ}}}%             % fractional degree symbol:     0.¡0

\begin{acronym}[AWGN]
\acro{2MASS}{Two Micron All Sky Survey}
\acro{2MASX}{Two Micron All Sky Survey Extended Source Catalogue}
\acro{30Dor}[30 Dor]{30~Doradus}
\acro{AeReS}{Aegean Residual}
\acro{AGN}{active galactic nucleus}
\acro{ALMA}{Atacama Large Millimeter Array}
\acro{ARC}{Australian Research Council}
\acro{ATCA}{Australia Telescope Compact Array}
\acro{ATESP}{Australia Telescope ESO Slice Project}
\acro{ATNF}{Australia Telescope National Facility}
\acro{ATOA}{Australia Telescope Online Archive}
\acro{AT20G}{Australia Telescope 20 GHz Survey}
\acro{ASKAP}{Australian Square Kilometre Array Pathfinder}
\acro{BETA}{Boolardy Engineering Test Array}
\acro{BGG}{Brightest Galaxy Group}
\acro{BL}[BL Lac]{BL Lacertae Objects}
\acro{CASS}{CSIRO Astronomy and Space Science}
\acro{CABB}{Compact Array Broadband Back-end}
\acro{CARTA}{Cube Analysis and Rendering Tool for Astronomy}
\acro{CASDA}{CSIRO ASKAP Science Data Archive}
\acro{CC}{core-collapse}
\acro{CHII}[{\sc CHii}]{Compact \textsc{Hii}}
\acro{CIE}{collisional ionisation equilibrium}
\acro{CMB}{Cosmic Microwave Background}
\acro{CSIRO}{Australian Commonwealth Scientific and Industrial Research Organisation}
\acro{CSS}{Compact Steep Spectrum}
\acro{CTA}{Cherenkov Telescope Array}
\acro{DESI}{Dark Energy Spectroscopic Instrument}
\acro{DS9}[\textsc{DS9}]{\textsc{SAOImage DS9}}
\acro{DSS}{Digital Sky Survey}
\acro{DSA}{diffuse shock acceleration}
\acro{EBHIS}{Effelsberg--Bonn H\,{\sc i} Survey}
\acro{EM}{Electromagnetic}
\acro{EMU}{Evolutionary Map of the Universe}
\acro{eRASS}{{\it eROSITA} All-Sky Survey}
\acro{ev}[eV]{electronvolt\acroextra{: 1 eV $\approx 1.6 \times 10^{-19}$ J}}
\acro{eROSITA}{extended R\"Ontgen Survey with an Imaging Telescope Array}
\acro{EVLA}{The Expanded Very Large Array}
\acro{FAST}{The Five-hundred-meter Aperture Spherical radio Telescope}
\acro{Fermi-LAT}{Fermi Large Area Telescope}
\acro{FIR}{Far Infrared}
\acro{FITS}[\textsc{Fits}]{Flexible Image Transport System}
\acro{FRB}{Fast Radio Bursts}
\acro{FRED}{fast-rise, exponential-decay}
\acro{FSRQ}{Flat Spectrum Radio Quasars}
\acro{FUSE}{Far Ultraviolet Spectroscopic Explorer}
\acro{FWHM}{Full Width at Half-Maximum}
\acro{GASS}{Galactic All-Sky Survey}
\acro{GLEAM}{GaLactic and Extragalactic All-sky MWA}
\acro{GLEAM-X}{GaLactic and Extragalactic All-sky MWA eXtended}
\acro{GPS}{Gigahertz Peak Spectrum}
\acro{HEMT}{High Electron Mobility Transistor}
\acro{HERITAGE}{HERschel Inventory of The Agents of Galaxy Evolution}
\acro{H.E.S.S.}{High Energy Stereoscopic System}
\acro{HFP}[HFP]{High-Frequency Peaker}
\acro{HPBW}{Half Power Beam Width} 
\acro{HzRGs}{High Redshift Radio Galaxies}
\acro{pHFP}[pHFP]{Potential High Frequency Peaker}
\acro{HI}[H{\sc i}]{Neutral Atomic Hydrogen} 
\acro{HST}{\textit{Hubble Space Telescope}}
\acro{ICM}{intracluster medium}
\acro{IAU}{International Astronomical Union}
\acro{IFRSs}{Infrared Faint Radio Sources}
\acro{ISM}{interstellar medium}
\acro{IFRS}{Infrared Faint Radio Source}
\acro{IR}{infrared}
\acro{JY}[Jy]{Jansky\acroextra{, 1 Jy = $10^{-26} \times \mathrm{W~ m}^{-2}~\mathrm{Hz}^{-1}$}} 
\acro{LOFAR}{LOw Frequency ARray}
\acro{LLAGN}{low-luminosity active galactic nuclei}
\acro{LINER}{Low-Ionisation Narrow Emission-Line Region}
\acro{LLS}{Largest Linear Size}
\acro{LFAA}{Low-Frequency Aperture Array}
\acro{LMC}{Large Magellanic Cloud}
\acro{LSO}[LSOs]{Large Scale Objects}
\acro{MACHO}{Massive Astrophysical Compact Halo Objects}
\acro{MAGMA}{Magellanic Mopra Assessment}
\acro{MC}[MC]{Magellanic Cloud}
\acro{MS}[MS]{Magellanic Stream} 
\acro{MB}[MB]{Magellanic Bridge} 
\acro{MCELS}{Magellanic Cloud Emission Line Survey}
\acro{MIR}{mid-infrared}
\acro{MSX}[{\it MSX}]{{\it Midcourse Space Experiment}}
\acro{MW}{Milky Way}
\acro{MIPS}{Multiband Imaging Photometer}
\acro{MIRIAD}[\textsc{Miriad}]{Multichannel Image Reconstruction, Image Analysis and Display}
\acro{MIT}{Massachusetts Institute of Technology}
\acro{MM}{mixed-morphology}
\acro{MOST}{Molonglo Observatory Synthesis Telescope}
\acro{MQS}{Magellanic Quasars Survey}
\acro{MRC}{Molonglo Reference Catalogue of Radio Sources}
\acro{MT-MFS}{Multi-Term Multi-Frequency Synthesis}
\acro{MWA}{Murchison Widefield Array}
\acro{NED}{NASA/IPAC Extragalactic Database}
\acro{NEI}{non-equilibrium ionisation}
\acro{NRAO}{National Radio Astronomy Observatory}
\acro{NVSS}{NRAO VLA Sky Survey}
\acro{OPAL}{Online Proposal Applications \& Links}
\acrodefplural{ORC}{Odd Radio Circles}
\acro{ORC}{Odd Radio Circle}
\acro{OVV}{Optically Violent Variable Quasars}
\acro{PACS}{Photodetector Array Camera and Spectrometer}
\acro{PAF}{Phased Array Feed}
\acro{pc}{parsec\acroextra{: 1 pc $\simeq 3.09 \times 10^{16}$ m}}
\acro{PMN}{Parkes-MIT-NRAO}
\acro{PN}{Planetary Nebula}
\acro{PWN}{pulsar wind nebula}
\acrodefplural{PWN}[PWNe]{pulsar wind nebulae}
\acro{RN}{reflection nebula}
\acrodefplural{RN}[RNe]{reflection nebulae}
\acro{QSO}{Quasi-Stellar Object}
\acro{RA}{Right Ascension}
\acro{RFI}{Radio-Frequency Interference}
\acro{RIAF}{radiatively inefficient accretion flow}
\acro{RMS}{Root Mean Squared}
\acro{RM}{rotation measure}
\acro{SAGE}{Surveying the Agents of a Galaxy’s Evolution}
\acro{SARAO}{South African Radio Astronomy Observatory}
\acro{SD}{single-degenerate}
\acro{SDSS}{Sloan Digital Sky Survey}
\acro{SED}{spectral energy distribution}
\acro{SI}[$\alpha$]{Spectral Index\acroextra{, $S \propto \nu^\alpha$}}
\acro{SHS}{SuperCOSMOS H$\alpha$ Survey}
\acro{SKA}{Square Kilometre Array}
\acro{SMASH}{Survey of the MAgellanic Stellar History}
\acro{SMBH}{supermassive black hole}
\acro{SMBBH}{super massive binary black hole}
\acro{POSSUM}{Polarisation Sky Survey of the Universe's Magnetism}
\acro{SMGPS}{SARAO MeerKAT Galactic Plane Survey}
\acro{SMC}{Small Magellanic Cloud}
\acro{SN}{Supernova}
\acro{SNR}{supernova remnant}
\acro{SPIRE}{Spectral and Photometric Imaging Receiver}
\acro{SUMSS}{Sydney University Molonglo Sky Survey}
\acro{TOPCAT}[\textsc{Topcat}]{Tool for OPerations on Catalogues And Tables}
\acro{TDE}{tidal disruption event}
\acro{USS}{Ultra Steep Spectrum}
\acro{UV}{ultra-violet}
\acro{YSOs}{young stellar objects}
\acro{WBAC}{Wide-Band Analogue Correlator}
\acro{WIFES}[WiFeS]{Wide-Field Spectrograph}
\acro{WISE}{Wide-Field Infrared Survey Explorer}
\acro{VLBI}{Very Long Baseline Interferometry} 
\acro{VLSR}[\textbf{$v_{lsr}$}]{Velocity in the Line of Sight}
\acro{WR}{Wolf-Rayet}
\acro{DR2}{Data Release 2}
\acro{DR3}{Data Release 3}
\acro{LBV}{Luminous Blue Variable}
\acro{RSG}{Red Supergiant}
\end{acronym}

\begin{document}
\parindent=.5cm
\baselineskip=3.8truemm
\columnsep=.5truecm
\newenvironment{lefteqnarray}{\arraycolsep=0pt\begin{eqnarray}}
{\end{eqnarray}\protect\aftergroup\ignorespaces}
\newenvironment{lefteqnarray*}{\arraycolsep=0pt\begin{eqnarray*}}
{\end{eqnarray*}\protect\aftergroup\ignorespaces}
\newenvironment{leftsubeqnarray}{\arraycolsep=0pt\begin{subeqnarray}}
{\end{subeqnarray}\protect\aftergroup\ignorespaces}
\def\arcsec{\hbox{$^{\prime\prime}$}}

% Runningtitle

\markboth{\eightrm Radio study of \cohen} 
{\eightrm M. D. FILIPOVI\'C {\lowercase{\eightit{et al.}}}}

\begin{strip}

{\ }

\vskip-1cm

\publ

\type

{\ }

% Title

\title{MeerKAT and ASKAP EMU new radio-continuum observations of Wolf-Rayet nebula PCG~11 (\cohen)}

% Authors

\authors{
M. D. Filipovi\'c$^1$,
Z. J. Smeaton$^1$,
A. C. Bradley$^1$,
R. Z. E. Alsaberi$^{1,2}$,
H. Sano$^{2, 3, 4, 5}$,
}
\authors{
L. Barnes$^1$,
C. Buemi$^6$,
C. Bordiu$^7$,
A. Brunetti$^{1,8}$,
F. Bufano$^6$,
F. Cavallaro$^6$,
}
\authors{
E. J. Crawford$^1$,
Y. Fukui$^{2, 9}$,
A. J. Green$^{10}$,
A. M. Hopkins$^{11}$, 
A. Ingallinera$^6$,
}
\authors{
R. Kothes$^{12}$,
S. Lazarevi\'c$^{1,13,14}$,
D. Leahy$^{15}$,
S. Loru$^6$,
K. Luken$^1$,
M. Mallows$^1$,
}
\authors{
Q. A. Parker$^{16}$,
J. L. Payne$^1$,
G. Rowell$^{17}$,
M. Stupar$^1$,
N. F. H. Tothill$^1$,
K. Tsuge$^2$,
}
\authors{
G. M. Umana$^6$,
D. Uro\v sevi\'c$^{18}$,
S. F. ur Rahman$^{19}$,
and J. Th. van Loon$^{20}$
}

\vskip3mm

% Address

\address{$^1$Western Sydney University, Locked Bag 1797, Penrith South DC, NSW 2751, Australia}
\address{$^2$Faculty of Engineering, Gifu University, 1-1 Yanagido, Gifu 501-1193, Japan}
\address{$^{3}$Center for Space Research and Utilization Promotion (c-SRUP), Gifu University, 1-1 Yanagido, Gifu 501-1193, Japan}
\address{$^{4}$Department of Intelligence Science and Engineering, Graduate School of Natural Science and Technology, Gifu University, 1-1 Yanagido, Gifu, 501-1193 Japan}
\address{$^{5}$Graduate School of Science, Nagoya University Southern Observatories, Nagoya University, Furo-cho, Chikusa-ku, Nagoya 464-8601, Japan}
\address{$^{6}$Instituto de Astrofísica de Andalucía (IAA-CSIC), Spain}
\address{$^{7}$INAF-Osservatorio Astrofisico di Catania, Via Santa Sofía 78, I-95123 Catania, Italy}
\address{$^{8}$Dipartimento di Fisica, Università di Torino, via P. Giuria 1, I–10125 Torino, Italy}
\address{$^{9}$ Department of Physics, Nagoya University, Furo-cho, Chikusa-ku, Nagoya 464-8601, Japan}
\address{$^{10}$Sydney Institute for Astronomy (SIfA), School of Physics, The University of Sydney, NSW 2006, Australia}
\address{$^{11}$School of Mathematical and Physical Sciences, 12 Wally’s Walk, Macquarie University, NSW 2109, Australia}
\address{$^{12}$Dominion Radio Astrophysical Observatory, Herzberg Astronomy \& Astrophysics, National Research Council Canada, P.O. Box 248, Penticton}
\address{$^{13}$Australia Telescope National Facility, CSIRO, Space and Astronomy, PO Box 76, Epping, NSW 1710, Australia}
\address{$^{14}$Astronomical Observatory, Volgina 7, 11060 Belgrade, Serbia}
\address{$^{15}$Department of Physics and Astronomy, University of Calgary, Calgary, Alberta, T2N IN4, Canada}
\address{$^{16}$Laboratory for Space Research, Faculty of Science, The University of Hong Kong, Hong Kong (SAR), People's Republic of China }
\address{$^{17}$School of Physics, Chemistry and Earth Sciences, The University of Adelaide, Adelaide, 5005, Australia}
\address{$^{18}$Department of Astronomy, Faculty of Mathematics, University of Belgrade, Studentski trg 16, 11000 Belgrade, Serbia}
\address{$^{19}$Lahore University of Management Sciences, Lahore, Pakistan}
\address{$^{20}$Lennard-Jones Laboratories, Keele University, ST5 5BG, UK}

% E-mail

\Email{M.Filipovic@westernsydney.edu.au}

\vskip3mm

% Received and Accepted dates

\dates{}{}

% Abstract

\summary{We present new radio-continuum observations of the well-known \ac{WR} nebula PCG~11 (\cohen; named after and in honour of the co-author on the discovery paper, Martin Cohen) using observations from the \ac{ASKAP} \ac{EMU}, \ac{ATCA} and MeerKAT surveys. We find that the elliptical shell surrounding \cohen\ has a flux density of 62$\pm$2\,mJy at 1000~MHz. The angular size of the elliptical emission is measured to be 54\arcsec$\times$45\arcsec\ (at P.A.=69\fdg4), which agrees well with observations of the source in H$\alpha$ and at 24\,$\mu$m in \ac{IR}. Using the measured angular size and a distance of 3.94$\pm$0.27\,kpc, we determine that the nebula has a linear size of $1.03(\pm0.07)\times0.86(\pm0.06)$\,pc.  
Together with our new and previously measured flux density estimates, we determine a spectral index for \cohen\ to be $\alpha$=0.06$\pm$0.02. %and for \cohenstar\ to be $\alpha$=1.05$\pm$0.24, 
indicating that the emission from the \ac{WR} nebula is of thermal origin. Based on a typical average expansion velocity of 75~km~s$^{-1}$ for a \ac{WR} star with a \ac{RSG} progenitor, we estimate the age of \cohen\ to be $\sim$ 13500~years. Our NANTEN telescope CO observations suggest that \cohen\ was formed during a relatively recent mass-loss event, while the associated CO-emitting molecular gas reflects the cumulative effects of stellar-wind feedback over a much longer timescale ($\sim$1.5\,Myr). As one of the few \ac{WR} ring nebulae with linked molecular gas, \cohen\ provides a valuable laboratory for studying the long-term impact of massive-star feedback on surrounding molecular clouds.} 

% Keywords (see keywords.pdf file)


\end{strip}

\tenrm

\acresetall

\section{INTRODUCTION}
\label{sec:1}
\indent
 
\ac{WR} stars are massive late-type stars with broad emission lines \citep{2007ARA&A..45..177C} and are easily identifiable through their mass-loss rates \citep{2000A&A...360..227N} and visually striking outbursts across the electromagnetic spectrum \citep{1995AJ....109.1839M}. These stars are characterised by the elements that dominate their spectra; the primary elemental emissions being carbon, nitrogen, or oxygen \citep{2007ARA&A..45..177C}. According to the dominant spectral emission, \ac{WR} spectral classifications are termed WC (carbon-rich), WN (nitrogen-rich), and WO (oxygen-rich). \ac{WR} stars have unique evolutionary cycles tied to their mass-loss history, with variable and inconsistent outbursts driven by internal pulsations \citep{1994A&A...290..819L}.

\ac{WR} stars have been known to emit at radio frequencies for some time \citep{1995ApJ...450..289L}, but these detections were often attributed to the stars themselves rather than their ejecta. Recent surveys at radio frequencies have revealed several previously uncharacterised \ac{WR} shells that are well known at other wavelengths (typically optical and \ac{IR}). Examples of these are WR16 and WR40 \citep{2025PASA...42..101B,2025SerAJ.211...65B}, WN8h stars with prominent emission at 944~MHz as seen with the \ac{ASKAP} \ac{EMU} survey (see below). Late WNh stars are preferentially detected in radio \citep{2021MNRAS.506.3540P,2026MNRAS.549ag963C}, as these ones also turn out to be. These are among the most massive stars, and actually do not descend from red supergiants. Here, we present new radio detections of the shell surrounding a known \ac{WR} star, \cohen.

Over the last several years, the population of known Galactic radio sources has grown substantially, driven largely by a handful of \ac{SKA} precursor surveys that are conducting surveys at better resolutions and sensitivities than previously possible. These include large-scale surveys undertaken by \ac{ASKAP}, such as the \ac{EMU} survey~\citep{Norris2021,2025PASA...42...71H}, MeerKAT, such as the \ac{SMGPS}~\citep{2024MNRAS.531..649G}, as well as by other precursors such as the \ac{LOFAR}, \ac{FAST} \citep{2019SCPMA..6259502J}, and \ac{EVLA}\footnote{\url{https://science.nrao.edu/facilities/vla/docs/manuals/oss2013B/intro/project)}}. These instruments have led to the discovery and characterisation of several different astrophysical objects, including \acp{SNR}~\citep{2022MNRAS.512..265F,2023AJ....166..149F,2025PASA...42..104F,2025PASA...42...21M,2025A&A...693L..15S,2026A&A...710A.308B,2026A&A...710A..84L} and \ac{SNR} candidates~\citep{2024MNRAS.534.2918S,2024RNAAS...8..158S,2024RNAAS...8..107L}, \acp{RN}~\citep{2025PASA...42...32B}, \acp{PWN}~\citep{2024PASA...41...32L,2026ApJ..1007..159L,2025MNRAS.537.2868A,2026PASA...43...69A}, and others~\citep{2022MNRAS.516.1865V,2023MNRAS.523.1933V,2024A&A...690A..53B,2025MNRAS.543.3708B,Filipovic_2025,2025PASA...42..119S,2026PASA...43...22K,2026PASA...43...35Z}. They have also been useful in enabling more detailed analyses of previously known radio objects~\citep{2024PASA...41..112F, 2025PASA...42.e111A, 2025SerAJ.210...13S, 2025SerAJ.211...27G, 2026arXiv260626889I, 2026arXiv260628210C,Filipovic2026EMU} due to their improved sensitivity and resolution. They were paramount in the discovery of the enigmatic class of radio objects known as \acp{ORC}~\citep{2021PASA...38....3N,2022MNRAS.513.1300N,2021MNRAS.505L..11K,2025A&A...702A.219T,2026PASA...43...43F} and have significantly improved the currently available set of radio catalogues~\citep{2021MNRAS.506.3540P,2023MNRAS.524.1396B,2025ApJ...988...75B,2024MNRAS.529.2443C,2026MNRAS.549ag963C,2026arXiv260305966R}. 

\cohenstar\ and \cohen, also known as PCG~11 \citep{2012MNRAS.427.3016P}, PHR~1633--4928 and WR75ab, is a known \ac{WR} star with an associated ring nebula. The star (\cohenstar) itself and the nebula (\cohen) were first discovered and classified by \citet{2005MNRAS.360.1439C} using H$\alpha$ data from the AAO/UKST H$\alpha$ Survey of the Southern Galactic Plane~\citep{1998A&G....39d..10P}. They confirmed the star's WN7h type classification with optical spectroscopy\footnote {Nitrogen-rich, subtype~7 with characteristic stellar surface temperature $\sim$50--60\,kK, having detectable hydrogen in its winds.}, determined a distance of 4.1$\pm$0.4\,kpc, and mapped the radio-continuum properties of the ring nebula using radio-continuum data from \ac{SUMSS} and \ac{MIR} data from \ac{MSX}. Following this, the central star has been analysed using \textit{Gaia} \ac{DR2} and \ac{DR3} data~\citep{2016A&A...595A...1G,2018A&A...616A...1G,2020MNRAS.493.1512R,2023A&A...674A...1G}, and the most recent distance estimate from \ac{DR3} is 3.94$\pm$0.27\,kpc~\citep{2025MNRAS.543.1158A}, which we adopt for the calculations discussed in Section~\ref{sec:3}. 
\cohen\ now appears in the MeerKAT radio surveys as well as in \ac{ATCA} observations \citep{2026arXiv261103954B}, providing a higher-resolution view of the nebula's radio-continuum morphology. 

Among Galactic \ac{WR} nebulae, \cohen\ stands out for its exceptionally symmetric elliptical morphology, visible across optical, infrared, and radio wavelengths. The availability of new \ac{ASKAP} \ac{EMU} and MeerKAT observations now provides the first opportunity to investigate the radio-continuum properties of the nebula with greatly improved sensitivity and angular resolution. Moreover, combining radio-continuum, optical, infrared, and molecular-gas observations allows us to explore massive-star feedback across multiple spatial and temporal scales. While the compact radio shell traces a relatively recent mass-loss episode from the central \ac{WR} star, the surrounding molecular-gas environment may preserve evidence of the cumulative impact of stellar winds over a much longer evolutionary timescale. As one of the relatively small number of \ac{WR} ring nebulae with an associated molecular structure, \cohen\ provides an important laboratory for investigating the interaction between massive stars and their surrounding interstellar medium.

Determining the fundamental physical properties of \cohen, particularly its distance, physical size and evolutionary timescale, is essential for establishing the nature and evolutionary state of the nebula. These parameters provide the physical framework required to interpret its morphology and multi-wavelength emission and to distinguish between possible formation and evolutionary scenarios. \cohen\ is particularly well suited to such an investigation because of its remarkably coherent, nearly circular morphology and its detection across multiple wavelength regimes, which enable complementary diagnostics to probe its structure and emission properties. Despite these favourable characteristics, several of its basic physical properties remain poorly constrained. In this work, we combine new radio-continuum observations with available multi-wavelength data to provide a more complete characterisation of \cohen\ and to place improved constraints on its physical dimensions, emission properties and evolutionary state.
In Section~\ref{sec:2}, we describe the observational data and analytical techniques used; in Section~\ref{sec:3}, we present the analysis results; and in Section~\ref{sec:4}, we present the conclusions.

\section{DATA}
\label{sec:2}

\subsection{ASKAP EMU}
\label{Section:DATA:ASKAP_EMU}
\indent

\begin{figure*}[!ht]
    \centering
    \includegraphics[width=\linewidth]{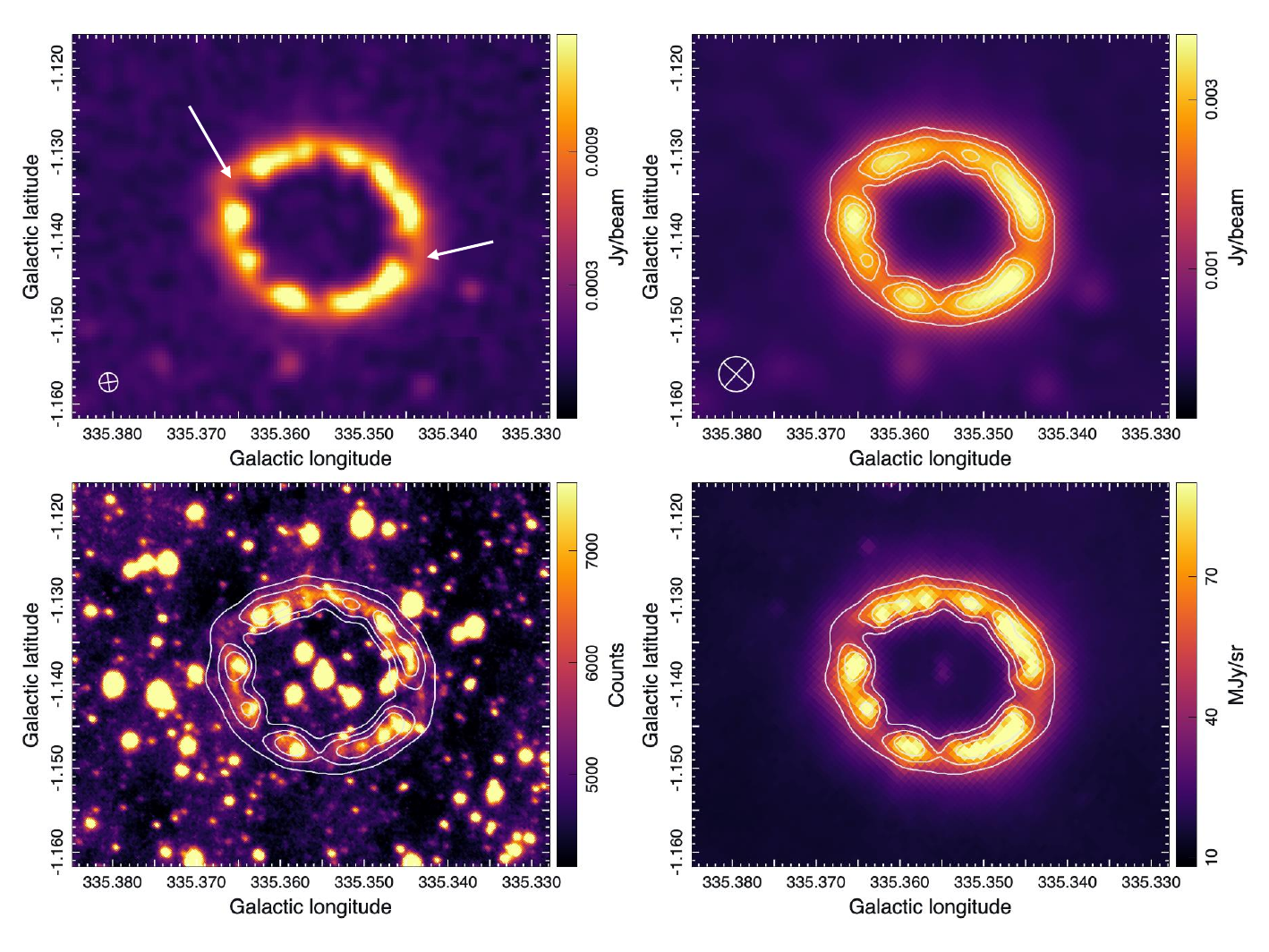}
    \caption{4-panel image showing \cohen\ at different frequencies. These are radio-continuum images from MeerKAT at 1300\,MHz (top-left) and \ac{ASKAP} at 944\,MHz (top-right), both showing the beam size in the bottom-left corners, the \Halpha\ optical image from SuperCOSMOS (bottom-left) and a 24$\mu$m IR image from the {\it Spitzer} telescope (bottom-right). All images are linearly scaled. The ASKAP, H$\alpha$, and {\it Spitzer} images have the MeerKAT radio-continuum contours overlaid at levels of 0.4, 0.8, and 1.2\,mJy\,beam$^{-1}$, corresponding to levels of approximately 7, 15, and 22$\sigma$ for an estimated MeerKAT RMS noise level of $\sim$55\,$\mu$Jy\,beam$^{-1}$. The arrows in the MeerKAT radio-continuum image (top-left) show the locations of the holes in the ring structure, as discussed in Section~\ref{subsec:morph}.} 
    \label{fig:4-pan}
\end{figure*}

\cohen\ and its shell have been detected in one \ac{EMU} tile (EMU$\_1638$-$51$), which was observed during the \ac{ASKAP} schedule block SB85620 on 14$^\text{th}$~June~2026. This observation was reduced using the standard \ac{ASKAP} pipeline, ASKAPSoft, using multi-frequency synthesis imaging, multi-scale cleaning, self-calibration and convolution to a common beam size of 15\arcsec$\times$15\arcsec~\citep{guzman2019}. \cohen\ is located close to Galactic Plane emission, placing it in an area of slightly higher background noise. We estimate a local \ac{RMS} noise level of $\sim$150\,$\mu$Jy\,beam$^{-1}$. The final image is shown in Figure~\ref{fig:4-pan} (top-right).

%%%%%%%%%%%%%%%%%%%%%%%%%%%%%%%%%%%%%%%%%%%%%%%%  FIG 1  %%%%%%%%%%%%%%%%%%%%%%%%%%%%%%%%%%%%%%%%%%%%%%%%%%%%%%%%%%%%%%%%%%%%%%%%%%%%%%%%%
%%%%%%%%%%%%%%%%%%%%%%%%%%%%%%%%%%%%%%%%%%%%%%%%%%%%%%%%%%%%%%%%%%%%%%%%%%%%%%%%%%%%%%%%%%%%%%%%%%%%%%%%%%%%%%%%%%%%%%%%%%%%%%%%%%%%%%%%%

\subsection{MeerKAT}
\label{Section:DATA:MeerKAT}
\indent

The MeerKAT observational data were obtained from the recent MeerKAT survey of the Galactic Plane, \ac{SMGPS}~\citep{2024MNRAS.531..649G}. The data consist of cubes over the MeerKAT L-Band of 908--1656\,MHz (centred at 1300\,MHz), power-law fits to the images, and broad-band images with an angular resolution of 8\arcsec$\times$8\arcsec\ and an average \ac{RMS} sensitivity of $\sim$10--20\,$\mu$Jy\,beam$^{-1}$. The survey extended to Galactic latitudes $|b| \leq 1\fdg5$. \cohen\ is located near the edge of this latitude limit, so the power-law fit images were unavailable. Therefore, we use only the broadband-averaged image and measure a local \ac{RMS} noise level of $\sim$55\,$\mu$Jy\,beam$^{-1}$ due to the slightly complex area, which results in higher noise. We use only the Stokes~$I$ broadband data, and further details on the data reduction are discussed by \citet{2024MNRAS.531..649G}. The final image is shown in Figure~\ref{fig:4-pan} (top-left).

\subsection{ATCA}
\label{Section:DATA:ATCA}
\indent
\ac{ATCA} archival\footnote{\ac{ATOA}, hosted by the \ac{ATNF}: \url{https://atoa.atnf.csiro.au}} data (project code: C3526) were used to produce high resolution images for \cohen\ and \cohenstar\ centred at 5.5 and 9\,GHz. The preliminary images used in this analysis had resolutions and measured local noise levels of 3\arcsec$\times$3\arcsec\ and $\sim$25\,$\mu$Jy\,beam$^{-1}$ for the 5.5\,GHz image and 2\arcsec$\times$2\arcsec\ and $\sim$20\,$\mu$Jy\,beam$^{-1}$ for the 9\,GHz image. These \ac{ATCA} data, along with further details on the imaging, observations, and in-depth analysis, will be presented in \citet{2026arXiv261103954B}.

\subsection{NANTEN}
\label{subsec:NANTEN}
To investigate the molecular gas environment of \cohen, we used an archival $^{12}$CO($J$~=~1--0) dataset obtained with the NANTEN 4-m millimeter radio telescope \citep{2004ASPC..317...59M}. The dataset has an angular resolution of 2\farcm6 and a velocity resolution of 0.65\,km\,s$^{-1}$, with a typical rms noise level of $\sim$0.3\,K.

\subsection{Optical (SuperCOSMOS) and IR (Spitzer)}
\label{Section:DATA:Optical_IR}
\indent

We also include observations from the SuperCOSMOS \Halpha\ survey~\citep[656.281~nm;][]{2005MNRAS.362..689P} and from the {\it Spitzer} space telescope \ac{MIPS} instrument~\citep[24~$\mu$m;][]{2004ApJS..154...25R}\footnote{{\it Spitzer} data was obtained from the public Spitzer Heritage Archive.}. We used these data to compare the nebulosity surrounding \cohen\ at different wavelengths. The final images are shown in Figure~\ref{fig:4-pan} (\Halpha\ in the bottom-left and {\it Spitzer} in the bottom-right).

\subsection{Other}
\label{Section:DATA:Other}
\indent

We checked several other surveys including the GaLactic and Extragalactic All-sky MWA ~\citep[GLEAM;][]{2015PASA...32...25W}, GaLactic and Extragalactic All-sky MWA eXtended~\citep[GLEAM-X;][]{2018PASA...35...33W,2022PASA...39...35H}, and {\it eROSITA} All-Sky Survey~\citep[eRASS;][]{2024A&A...682A..34M}\acused{eRASS}, but found no corresponding detections at the position of \cohen.

%%%%%%%%%%%%%%%%%%%%%%%%%%%%%%%%%%%%%%%%%%%%%%%%%%  FIG 2  %%%%%%%%%%%%%%%%%%%%%%%%%%%%%%%%%%%%%%%%%%%%%%%%%%%%%%%%%%%%%%%%%%%%%%%%%%%%%%
\begin{figure*}[!ht]
    \centering
    \includegraphics[width=\linewidth]{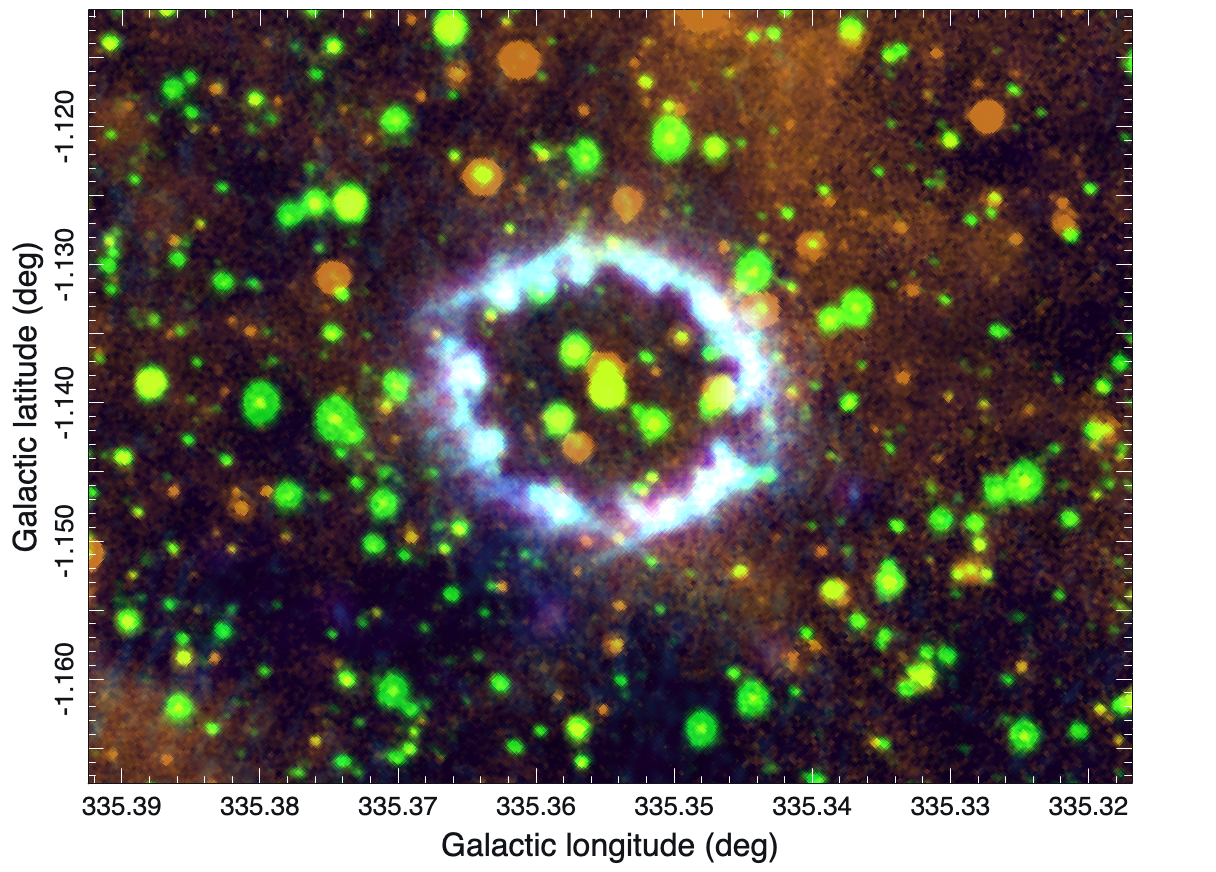}
    \caption{Multi-frequency composite colour image of \cohen. Purple is MeerKAT 1300\,MHz, cyan is \ac{ATCA} 5500\,MHz, orange is infrared IRAC~4 band, and green is optical \Ha.} 
    \label{fig:RGB}
\end{figure*}
%%%%%%%%%%%%%%%%%%%%%%%%%%%%%%%%%%%%%%%%%%%%%%%%%%%%%%%%%%%%%%%%%%%%%%%%%%%%%%%%%%%%%%%%%%%%%%%%%%%%%%%%%%%%%%%%%%%%%%%%%%%%%%%%%%%%%%%%%

\section{RESULTS AND DISCUSSION}
\label{sec:3}

Our focus in this study is on \cohen\ while \cohenstar\ in-depth analysis will be shown in \citet{2026arXiv261103954B}.
 
\subsection{Morphology and Dimensions of \cohen}
\label{subsec:morph}
\indent

Although different in nature (and origin), \cohen\ resembles the morphological appearance of the famous SN(R)\,1987A.
However, \cohen\ may be a merger product (common among massive stars) which might explain the ring, while the progenitor of SN(R)\,1987A is also explained as a merger product.

\cohen\ is centred (geometrical) at RA(J2000)=16$^{\rm h}$33$^{\rm m}$48\farcs7, DEC(J2000)=--49\D28\arcmin42\farcs9 ($l$=335\fdg35, $b$=--1\fdg14). We measure the angular size of \cohen\ from the MeerKAT radio-continuum image using the \ac{CARTA} \citep{2015ASPC..495..121R}, determining angular semi-axes of 54\arcsec$\times$45\arcsec\ (P.A.=69\fdg4 measured counter-clockwise from North in the Galactic coordinate system). \citet{2026arXiv261103954B} show an inclination (with respect to the sky) of 33\D, i.e. 57\D\ with respect to the line of sight, which is far from being perpendicular.

Using the distance from \citet{2005MNRAS.360.1439C} of 3.94$\pm$0.27\,kpc, we determine the linear size of the radio emission related to \cohen\ to be $1.03(\pm0.07)\times0.86(\pm0.06)$\,pc. 

The egg-like (elliptical) shape of the nebula (\cohen), detected in the new radio-continuum images, clearly resembles the H$\alpha$ and \ac{IR} morphology (Figure~\ref{fig:RGB}). We also detect the fractured rim structure reported by \citet{2005MNRAS.360.1439C}, shown by the annotations in Figure~\ref{fig:4-pan} (top-left). However, this is less prominent than in the \Halpha\ image. 

We tentatively detect \cohenstar\ only in the \ac{ATCA} radio observations\footnote{Centred at RA(J2000)=16$^{\rm h}$33$^{\rm m}$48.73$^{\rm s}$, DEC(J2000)=--49\D28\arcmin44\farcs0 which is $\sim$1\farcs1 south from the estimated geometrical centre of \cohen.}. 
%On the other side, \cohen\ appears clearly across all investigated wavebands (see Figure~\ref{fig:RGB}). 
An in-depth analysis of the \ac{ATCA} data will be presented in \citet{2026arXiv261103954B}, where they showed that the star appears to be brighter in the 9\,GHz image than in the 5.5\,GHz image, indicating an inverted spectrum, which would explain why the star is not visible in the lower-frequency radio-continuum observations \citep{2026arXiv261103954B}. 

Stars of similar type (WN7h) have been routinely detected between 5000-10000~MHz frequency range \citep{1995ApJ...450..289L,2024A&A...690A..78B}; however, it is important to note that the overwhelming majority of these detections occur at distances less than 3\,kpc \citep{1997ApJ...481..898L}. For comparison, WR82 is a \ac{WR} star situated at 5.25\,kpc, and has a reported non-detection limit of $<$0.23\,mJy at 8460\,MHz \citep{2004AJ....127.2885C}. It is also important to remember that \cohen\ is at Galactic Latitude $b=-$1\fdg14 in a particularly radio-crowded portion of the Galactic Plane. PCG~11 may be emitting at the \ac{EMU} frequency of 943.5~MHz, but because of its distance and the radio-bright Galactic Plane, we do not see it.

We attempt to put constraints on the age of \cohen\ by using canonical literature values of parameters of \ac{WR} stars, of which expansion velocities are the most important. A \ac{WR} star has a typical average expansion velocity between 50--600\,km\,s$^{-1}$, which is dependent on whether the star's previous life stage was as a \ac{LBV} or a \ac{RSG} \citep{2020MNRAS.495..417C}. A \ac{LBV} progenitor will have typically higher expansion velocities \citep[$\sim$100--600\,km\,s$^{-1}$;][]{2011IAUS..272..372W,2014MNRAS.442.3316S}, whereas a \ac{RSG} progenitor will be slightly lower \citep[$\sim$50-100\,km\,s$^{-1}$;][]{2000AJ....120.2670G}. When PCG~11 was first identified, \citet{2005MNRAS.360.1439C} classified it as a former \ac{RSG}, so its average expansion velocity is likely to be at the lower end of the discussed range. We assume an average expansion velocity of 75\,km\,s$^{-1}$; given that PCG~11 has semi-axes of 1.03\,pc (outer bound) and a distance of 3.94\,kpc, we estimate an age of $\sim$13500~years. Although it is important to note that this is approximate, and that the shell may be younger due to the star's canonical terminal wind velocity ($v_{\infty}$), which is 1300\,km\,s$^{-1}$ for a \ac{WR} star of WN7 type \citep{2007ARA&A..45..177C}, this age estimate is fairly consistent with similar \ac{WR} nebulae \citep{2025PASA...42..101B}.

We also considered whether existing \HI\ observations could provide an independent constraint on the expansion velocity. The HI4PI survey \citep{2016A&A...594A.116H} has sufficient spectral resolution to investigate the \HI\ velocity structure towards \cohen; however, its angular resolution of $\sim16\farcm2$ is too coarse to resolve the \HI\ distribution across the nebula adequately. Moreover, the location of \cohen\ close to the Galactic Plane introduces substantial confusion from unrelated Galactic \HI\ emission along the same line of sight. Consequently, even if multiple velocity components are present in the \HI\ spectrum, it is not possible to associate them unambiguously with the approaching and receding sides of \cohen. We therefore do not use the HI4PI data to derive or further constrain the expansion velocity.

\subsection{Flux Densities and Spectral Index}
\label{Section:DISCUSSION:Flux}

The flux density of the emission was previously measured from \ac{SUMSS} data as $S_{843\,\rm{MHz}}=47$\,mJy~\citep{2005MNRAS.360.1439C}. We measure accompanying flux densities from available radio data using the same region defined in Section~\ref{subsec:morph}. 
Before measuring ring flux densities, we smoothed all radio images to a common resolution of 15\arcsec$\times$15\arcsec. 
These values are obtained using \ac{CARTA}, assuming errors of up to 10\% as discussed in \citet{2022MNRAS.512..265F,2026PASA...43...43F}.

%%%%%%%%%%%%%%%%%%%%%%%%%%%%%%%%%%%%  TABLE 1 %%%%%%%%%%%%%%%%%%%%%%%%%%%%%%%%%%%%%%%%%%%%%%%%%%%%%%%%%%%%%%%%%%%%%%%%%%%%%%%%%%%%%%%%%
\begin{table*}
    \centering
    \small
    \begin{tabular}{|l|c|c|l|}
    \hline
Telescope   &  Frequency Density   & Flux       &  Reference \\
            &    (MHz)             & (mJy)         & \\
\hline
  SUMSS     & 843       & 47            & \citet{2005MNRAS.360.1439C} \\
  ASKAP     & 943       & 60$\pm$3  & This work \\
  MeerKAT   & 1280      & 66$\pm$3  & \citet{2026arXiv261103954B} \\
  MeerKAT   & 1300      & 64$\pm$3  & This work \\
  MeerKAT   & 2540      & 67$\pm$3  & \citet{2026arXiv261103954B} \\
  MeerKAT   & 3060      & 67$\pm$3  & \citet{2026arXiv261103954B} \\
  ATCA      & 5500      & 68$\pm$3  & This work \\
  ATCA      & 5500      & 70$\pm$4  & \citet{2026arXiv261103954B} \\
  ATCA      & 9000      & 68$\pm$3  & This work \\
  ATCA      & 9000      & 69$\pm$4  & \citet{2026arXiv261103954B} \\
\hline
$\alpha\pm\Delta\alpha$ &   ---       & 0.06$\pm$0.02 & This work \\
        \hline 
        \hline
    \end{tabular}
    \caption{Background-subtracted integrated flux densities and spectral index for the \cohen. No uncertainty was reported for the 843\,MHz value, so we assumed up to 10\% error in the fit in Figure~\ref{fig:specindexplot}.}
    \label{tab:Flux-Spectral}
\end{table*}
%%%%%%%%%%%%%%%%%%%%%%%%%%%%%%%%%%%%%%%%%%%%%%%%%%%%%%%%%%%%%%%%%%%%%%%%%%%%%%%%%%%%%%%%%%%%%%%%%%%%%%%%%%%%%%%%%%%%%%%%%%%%%%%%%%%%%%%%%

%%%%%%%%%%%%%%%%%%%%%%%%%%%%%%%%%%%%  FIG 3  %%%%%%%%%%%%%%%%%%%%%%%%%%%%%%%%%%%%%%%%%%%%%%%%%%%%%%%%%%%%%%%%%%%%%%%%%%%%%%%%%%%%%%%%%%
\begin{figure}[ht]
    \centering
    \includegraphics[width=1.1\linewidth]{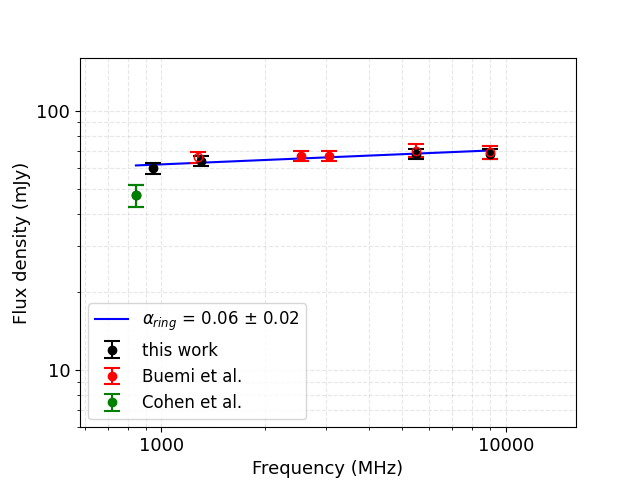}
    \caption{Spectral index plot for \cohen\ using the flux densities as shown in Table~\ref{tab:Flux-Spectral}. Uncertainties were not reported for the 843\,MHz point, so we have assumed up to a 10\% error in the fitting.} 
    \label{fig:specindexplot}
\end{figure}
%%%%%%%%%%%%%%%%%%%%%%%%%%%%%%%%%%%%%%%%%%%%%%%%%%%%%%%%%%%%%%%%%%%%%%%%%%%%%%%%%%%%%%%%%%%%%%%%%%%%%%%%%%%%%%%%%%%%%%%%%%%%%%%%%%%%%%%%%

%%%%%%%%%%%%%%%%%%%%%%%%%%%%%%%%%%%%%%%%%  FIG 4  %%%%%%%%%%%%%%%%%%%%%%%%%%%%%%%%%%%%%%%%%%%%%%%%%%%%%%%%%%%%%%%%%%%%%%%%%%%%%%%%%%%
\begin{figure}[ht]
    \centering
    \includegraphics[width=\linewidth]{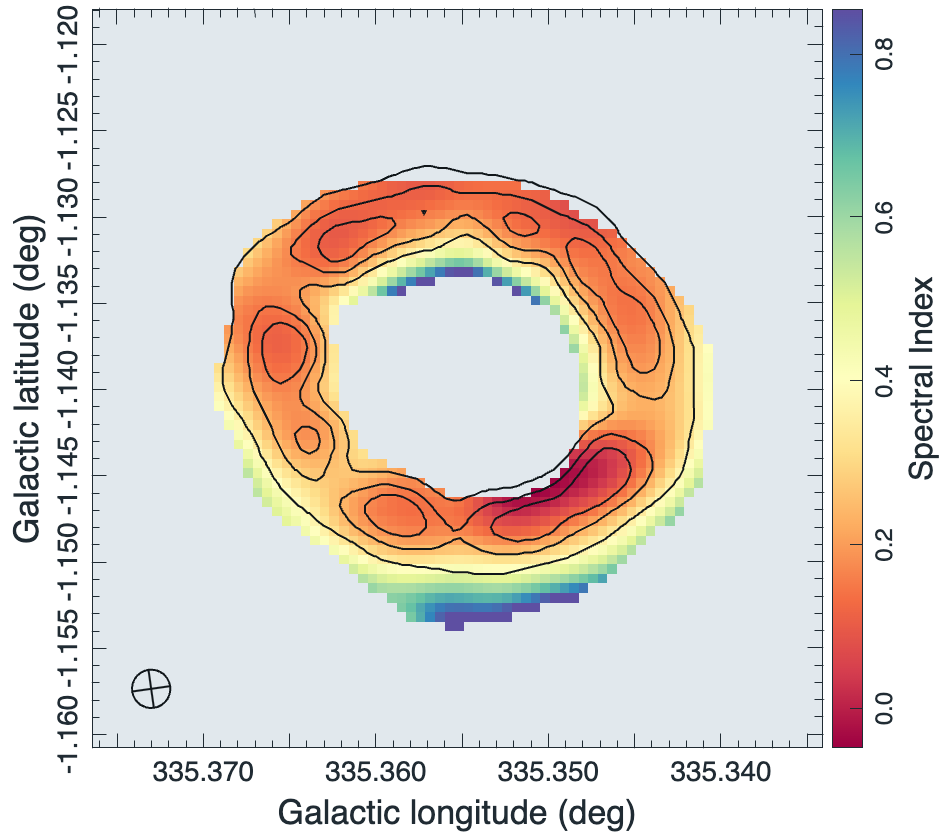}
    \caption{Spectral index map of \cohen\ generated using all available radio-continuum images. The final resolution is the lower \ac{ASKAP} value of 15\arcsec$\times$15\arcsec\ and the beam size is shown in the bottom left. The contours are from the MeerKAT radio-continuum image at levels of 0.4, 0.8, and 1.2\,mJy\,beam$^{-1}$.}
    \label{fig:specindex}
\end{figure}
%%%%%%%%%%%%%%%%%%%%%%%%%%%%%%%%%%%%%%%%%%%%%%%%%%%%%%%%%%%%%%%%%%%%%%%%%%%%%%%%%%%%%%%%%%%%%%%%%%%%%%%%%%%%%%%%%%%%%%%%%%%%%%%%%%%%%%%%%

The spectral index, $\alpha$, was calculated by fitting the measured flux densities ($S_{\nu}$) at different frequencies ($\nu$) using the standard relation $S\propto\nu^\alpha$ (Table~\ref{tab:Flux-Spectral}). 
A spectral index steeper than approximately $-0.3$ indicates non-thermal emission, while a flat or positive value suggests a thermal origin \citep{book2,book1}. 

For \cohen, we measure a spectral index of $\alpha=0.06\pm0.02$, implying a thermal origin for the emission from the \ac{WR} nebula.
We note that the data are consistent with a flat spectrum, except for the lowest-frequency point at 843\,MHz, which lies below the trend of the other measurements.
Therefore, the \cohen\ radio-continuum spectrum might consist of two theoretical slopes: an initial blackbody at low radio frequencies (optically thick environment) and the usual bremsstrahlung (optically thin), with a flat spectrum around $\alpha\sim0$ \citep{2015ApJ...805..119O}.  

Interpolating the best-fit relation from the data in Table~\ref{tab:Flux-Spectral}, we find a flux density at 1000\,MHz of 62$\pm$3\,mJy. We use this value to estimate the radio luminosity at 1000\,MHz using the relation $L_{1000\,{\textrm {MHz}}}\,=\,4\pi D^2_L S_{1000\,{\text{MHz}}}$ where $L_{1000\,{\textrm{MHz}}}$ is the radio luminosity at 1000\,MHz, $D_L$ is the distance to the object, taken as 3.94$\pm$0.27\,kpc, and $S_{1000\,{\textrm {MHz}}}$ is taken as 62$\pm$3\,mJy. We calculate a value of $L_{1000\,{\textrm {MHz}}}\,=\,(1.2\pm0.2)\times10^{14}$\,W\,Hz$^{-1}$.

To further constrain the spectral index and determine whether it varies across the source, we generate a high-resolution spectral index map using all available radio images. This was done using the \textsc{maths} function in the \textsc{miriad} software package. First, we convolved both images to the lowest resolution, in this case that of the \ac{ASKAP} image, B.S.=15\arcsec$\times$15\arcsec\ and P.S.=2\arcsec$\times$2\arcsec. %Both images were then cut at levels of 5$\sigma$ above the \ac{RMS}, and combined using the spectral index relationship $\alpha=(\log(S_{\rm MeerKAT})-\log(S_{\rm ASKAP}))/(\nu_{\rm MeerKAT}-\nu_{\rm ASKAP})$. 
We then combined the images to calculate the spectral index for each pixel in the map, following the relation above. The final spectral index map is shown in Figure~\ref{fig:specindex}, and we measure an average value of $\alpha\sim0.23$ over the whole ring structure, similar to results from the integrated flux density estimates. As shown in Figure~\ref{fig:specindex}, there is no obvious variation in the spectral index at present. 

This spectral index value is consistent with thermal emission, as there is no current evidence of synchrotron emission. The spectral shape may be more complex, but the current data cannot determine this. 

The radio-continuum properties of Cohen's Ring appear broadly consistent with those of other \ac{WR} nebulae recently studied with modern radio surveys. Similar thermal radio spectra have been reported for WR16 and WR40, where ionised circumstellar material associated with stellar mass loss dominates the radio emission. In this context, the measured spectral index and radio luminosity place \cohen\ comfortably within the thermal WR-nebula population. However, \cohen\ distinguishes itself through its remarkably regular elliptical morphology, which remains well preserved across radio, optical, and infrared wavelengths. While many evolved \ac{WR} nebulae exhibit substantial fragmentation, asymmetries, or evidence for strong interaction with an inhomogeneous interstellar medium, \cohen\ retains a highly symmetric shell structure. This suggests either a comparatively uniform local environment or a relatively early stage in the nebula's dynamical evolution, making it an excellent target for investigating the relationship between stellar mass loss and the surrounding interstellar medium.

\subsection{\cohen's Environment}
\label{subsec: environment}

\begin{figure}[]
    \centering
    \includegraphics[width=\columnwidth]{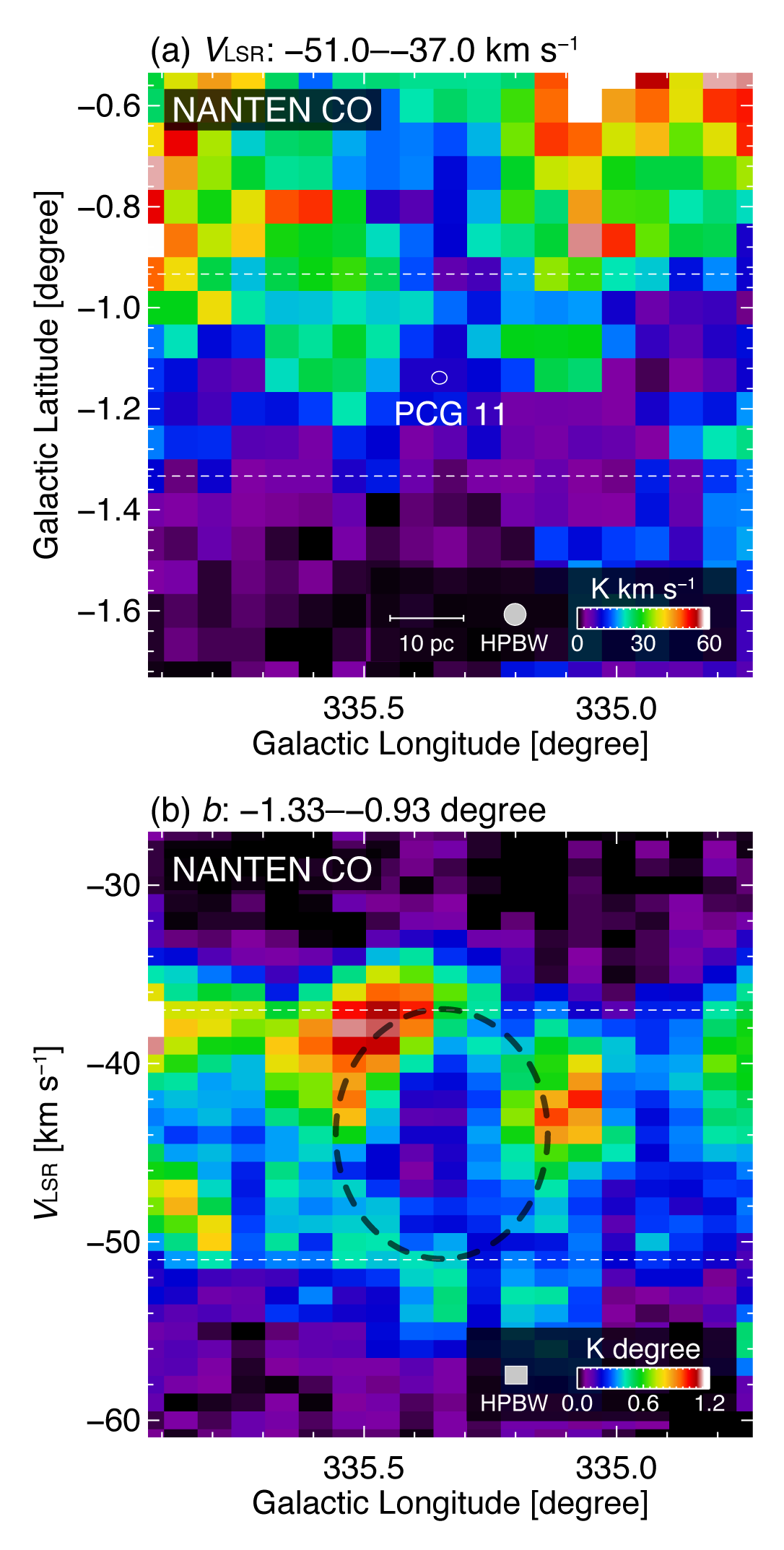}
    \caption{(a) Intensity distribution of the NANTEN $^{12}$CO($J$~=~1--0) line emission toward \cohen. The integration velocity range is from $-$51.0 to $-37.0$\,km\,s$^{-1}$. The ellipse shows the boundary of \cohen. (b) Position-velocity diagram of CO. The integration range of Galactic Latitude is from $-$1\fdg33 to $-$0\fdg93. The dashed semicircle corresponds to an expanding gas motion of CO likely due to the stellar wind from \cohen\ (see the text).} 
    \label{fig:NANTEN}
\end{figure}

Figure~\ref{fig:NANTEN}(a) shows the integrated intensity map of CO toward \cohen. We newly identified an incomplete, hollowed-out CO structure with a radius $R$ of approximately 0\fdg15, corresponding to $\sim$10\,pc at the adopted distance of 3.94\,kpc. \cohen, indicated by the ellipse, is located approximately at the geometric centre of this structure. This spatial correspondence suggests that the hollowed-out CO structure may represent a wind-blown bubble created by the stellar winds from the \ac{WR} star.

%%%%%%%%%%%%%%%%%%%%%%%%%%%%%%%%%%%%  FIG 5  %%%%%%%%%%%%%%%%%%%%%%%%%%%%%%%%%%%%%%%%%%%%%%%%%%%%%%%%%%%%%%%%%%%%%%%%%%%%%%%%%%%%%%%%%%
%%%%%%%%%%%%%%%%%%%%%%%%%%%%%%%%%%%%%%%%%%%%%%%%%%%%%%%%%%%%%%%%%%%%%%%%%%%%%%%%%%%%%%%%%%%%%%%%%%%%%%%%%%%%%%%%%%%%%%%%%%%%%%%%%%%%%%%%%

Figure~\ref{fig:NANTEN}(b) shows the position--velocity diagram of the same region. A cavity-like structure with a radius of $\sim$10~pc is seen over the velocity range from $V_\mathrm{LSR} \sim$$-51$ to $-37$\,km\,s$^{-1}$. This kinematic structure can be naturally interpreted as an expanding motion driven by the massive star's stellar winds. Under this interpretation, the expansion velocity is estimated to be $\Delta V \sim$$7$\,km\,s$^{-1}$, which is comparable to those observed in the molecular gas surrounding other \ac{WR} stars \citep[e.g.,][]{2025PASA...42..101B}.

We argue that the dynamical timescale of this larger expanding CO structure can be naturally explained if the expansion was driven by \cohenstar. The dynamical timescale of the CO shell $ t_\mathrm{dyn}$ is estimated as $t_\mathrm{dyn} \sim R/\Delta V \sim 10\,\mathrm{pc} / 7\;\mathrm{km\,s^{-1}} \sim 1.5\,\mathrm{Myr}$. This timescale broadly matches the duration over which a massive star can inject mechanical energy into its surroundings through stellar winds. Note that the CO expansion speed sets a lower limit on the speed at which the cavity was created (given the star's wind was faster), so $\sim 1.5\,\mathrm{Myr}$ is an upper limit on the duration of the cavity-blowing episode.

Taken together, these results suggest that \cohen, as traced by radio-continuum and optical emission, formed during a relatively recent mass-loss episode, whereas the expanding molecular gas traced by CO may record the cumulative effect of stellar-wind feedback over a much longer timescale of $\sim$1.5~Myr. Observational examples of molecular gas associated with ring nebulae harbouring \ac{WR} stars remain relatively limited. \cohen\ may therefore provide an important laboratory for investigating the long-term impact of stellar feedback from massive stars on the surrounding molecular clouds.

\section{CONCLUSION}
\label{sec:4}
\indent

We present new observations of the well-known \ac{WR} \cohen\ using observations from the \ac{ASKAP}, \ac{EMU}, \ac{ATCA} and MeerKAT surveys. The shell's elliptical shape is measured to have angular semi-axes of 54\arcsec$\times$45\arcsec, corresponding to a linear size of $1.03(\pm0.07)\times0.86(\pm0.06)$\,pc at a distance of 3.94$\pm$0.27\,kpc. We measure a flux density for \cohen\ at 1000~MHz to be 61$\pm$2\,mJy. We also calculate a spectral index of $\alpha=0.06\pm0.02$ (for \cohen), indicating that the emission from the \ac{WR} nebula is of thermal origin and at a possible age of $\sim$13500~years, assuming an expansion velocity of 75\,km\,s$^{-1}$.

We suggest that \cohen\ traces both a relatively recent mass-loss episode, as revealed by the radio-continuum and optical emission, and the longer-term influence of stellar-wind feedback, as reflected in the expanding CO-emitting molecular gas over a timescale of $\sim$1.5\,Myr. Given the rarity of molecular-gas detections associated with \ac{WR} ring nebulae, \cohen\ represents an important laboratory for investigating how massive stars shape and influence their surrounding molecular environment throughout their evolution.

% Acknowledgements
\section{ACKNOWLEDGEMENTS}

%% Authors %%
With the blessings of the surviving authors of the discovery paper \citep{2005MNRAS.360.1439C}, we rename this source as "\cohen" in honour and memory of our dear colleague and friend \fbox{Martin Cohen}. 

%% ASKAP %%
This scientific work uses data obtained from Inyarrimanha Ilgari Bundara / the Murchison Radio-astronomy Observatory. We acknowledge the Wajarri Yamaji People as the Traditional Owners and native title holders of the Observatory site. CSIRO’s ASKAP radio telescope is part of the Australia Telescope National Facility\footnote{\label{foot:ATNF}\url{http://www.atnf.csiro.au}}. Operation of ASKAP is funded by the Australian Government with support from the National Collaborative Research Infrastructure Strategy. ASKAP uses the resources of the Pawsey Supercomputing Research Centre. Establishment of ASKAP, Inyarrimanha Ilgari Bundara, the CSIRO Murchison Radio-astronomy Observatory and the Pawsey Supercomputing Research Centre are initiatives of the Australian Government, with support from the Government of Western Australia and the Science and Industry Endowment Fund.

%% MeerKAT %%
The MeerKAT telescope is operated by the South African Radio Astronomy Observatory, which is a facility of the National Research Foundation, an agency of the Department of Science and Innovation.

%% NANTEN %%
The NANTEN project is based on a mutual agreement between Nagoya University and the Carnegie Institution of Washington (CIW). YF greatly appreciates the hospitality of all staff members at CIW's Las Campanas Observatory. YF, HS, and KT are thankful to many Japanese public donors and companies who contributed to the realization of the project.

This research was supported by the Commonwealth through an Australian Government Research Training Program Scholarship [DOI:\url{https://doi.org/10.82133/C42F-K220}].

DU acknowledges the financial support provided by the Ministry of Science, Technological Development and Innovation of the Republic of Serbia through contract 451-03-136/2025-03/200104 and support through the joint project of the Serbian Academy of Sciences and Arts and Bulgarian Academy of Sciences ``Optical search for Galactic and extragalactic supernova remnants''. BA additionally acknowledges the funding provided by the Science Fund of the Republic of Serbia through project \#7337 ``Modeling Binary Systems That End in Stellar Mergers and Give Rise to Gravitational Waves'' (MOBY).

HS acknowledges financial support from the Japan Society for the Promotion of Science (JSPS) KAKENHI Grant Numbers JP20KK0309, JP24H00246, and JP26K21722.

We thank the referee for excellent comments and suggestions that greatly improved the manuscript.

% References

%\vskip2mm

\newcommand\eprint{in press }

\bibsep=0pt

\bibliographystyle{aa_url_saj}

{\small

\bibliography{PCG11}
}

\begin{strip}

\end{strip}

\clearpage

{\ }

\clearpage

{\ }

%\newpage

\begin{strip}
%\onecolumn

%{\ }

% Serbian abstract

% Title

\naslov{MeerKAT i ASKAP EMU nova radio-kontinuum posmatra{\nj}a Volf-Rajet nebule PCG~11 (Koenov Prsten)}

\authors{
{\rbf M. D. Filipovi{\cc}$^1$},
Z. J. Smeaton$^1$,
A. C. Bradley$^1$,
R. Z. E. Alsaberi$^{1,2}$,
H. Sano$^{2, 3, 4, 5}$,
}
\authors{
L. Barnes$^1$,
C. Buemi$^6$,
C. Bordiu$^7$,
A. Brunetti$^{1,8}$,
F. Bufano$^6$,
F. Cavallaro$^6$,
}
\authors{
E. J. Crawford$^1$,
Y. Fukui$^{2, 9}$,
A. J. Green$^{10}$,
A. M. Hopkins$^{11}$, 
A. Ingallinera$^6$,
}
\authors{
R. Kothes$^{12}$,
{\rbf S. Lazarevi{\cc}$^{1,13,14}$},
D. Leahy$^{15}$,
S. Loru$^6$,
K. Luken$^1$,
M. Mallows$^1$,
}
\authors{
Q. A. Parker$^{16}$,
J. L. Payne$^1$,
G. Rowell$^{17}$,
M. Stupar$^1$,
N. F. H. Tothill$^1$,
K. Tsuge$^2$,
}
\authors{
G. M. Umana$^6$,
{\rbf D. Uro{\ss}evi{\cc}$^{18}$},
S. F. ur Rahman$^{19}$,
and J. Th. van Loon$^{20}$
}

\vskip3mm

% Address

\address{$^1$Western Sydney University, Locked Bag 1797, Penrith South DC, NSW 2751, Australia}
\address{$^2$Faculty of Engineering, Gifu University, 1-1 Yanagido, Gifu 501-1193, Japan}
\address{$^{3}$Center for Space Research and Utilization Promotion (c-SRUP), Gifu University, 1-1 Yanagido, Gifu 501-1193, Japan}
\address{$^{4}$Department of Intelligence Science and Engineering, Graduate School of Natural Science and Technology, Gifu University, 1-1 Yanagido, Gifu, 501-1193 Japan}
\address{$^{5}$Graduate School of Science, Nagoya University Southern Observatories, Nagoya University, Furo-cho, Chikusa-ku, Nagoya 464-8601, Japan}
\address{$^{6}$Instituto de Astrofísica de Andalucía (IAA-CSIC), Spain}
\address{$^{7}$INAF-Osservatorio Astrofisico di Catania, Via Santa Sofía 78, I-95123 Catania, Italy}
\address{$^{8}$Dipartimento di Fisica, Università di Torino, via P. Giuria 1, I–10125 Torino, Italy}
\address{$^{9}$ Department of Physics, Nagoya University, Furo-cho, Chikusa-ku, Nagoya 464-8601, Japan}
\address{$^{10}$Sydney Institute for Astronomy (SIfA), School of Physics, The University of Sydney, NSW 2006, Australia}
\address{$^{11}$School of Mathematical and Physical Sciences, 12 Wally’s Walk, Macquarie University, NSW 2109, Australia}
\address{$^{12}$Australia Telescope National Facility, CSIRO, Space and Astronomy, PO Box 76, Epping, NSW 1710, Australia}
\address{$^{13}$Dominion Radio Astrophysical Observatory, Herzberg Astronomy \& Astrophysics, National Research Council Canada, P.O. Box 248, Penticton}
\address{$^{14}$Astronomical Observatory, Volgina 7, 11060 Belgrade, Serbia}
\address{$^{15}$Department of Physics and Astronomy, University of Calgary, Calgary, Alberta, T2N IN4, Canada}
\address{$^{16}$Laboratory for Space Research, Faculty of Science, The University of Hong Kong, Hong Kong (SAR), People's Republic of China }
\address{$^{17}$School of Physics, Chemistry and Earth Sciences, The University of Adelaide, Adelaide, 5005, Australia}
\address{$^{18}$Department of Astronomy, Faculty of Mathematics, University of Belgrade, Studentski trg 16, 11000 Belgrade, Serbia}
\address{$^{19}$Lahore University of Management Sciences, Lahore, Pakistan}
\address{$^{20}$Lennard-Jones Laboratories, Keele University, ST5 5BG, UK}

\Email{M.Filipovic@westernsydney.edu.au}

\vskip3mm

% UDC

\centerline{{\rrm UDK} \udc}

% Papertype

\vskip1mm

\centerline{\rit Uredjivaqki prilog}

\vskip.7cm

\baselineskip=3.8truemm

%\end{onecolumn}
\end{strip}

\clearpage

{\ }

%\clearpage

%\begin{strip}

%\begin{multicols}{1}

%\begin{onecolumn}

{
\rrm

Predstav{\lj}amo nova posmatra{\nj}a u radio-kontinumu dobro poznate Volf-Rajet magline PCG~11 (Koenov Prsten; nazvane po Martinu Koenu, koautoru rada u kojem je objekat otkriven, i {\nj}emu u qast), koriste{\cc}i posmatra{\nj}a iz \ac{ASKAP} \textrm{EMU} (Evoluciona Mapa Univerzuma), Australijskog Teleskopog Kompaktnog Niza (ATKN) i MeerKAT pregleda. 
Utvrdili smo da eliptiqna {\lj}uska koja okru{\zz}uje Koenov Prsten ima gustinu fluksa od 62$\pm$2\,\textrm{mJy} na 1000\,\textrm{MHz}. 
Ugaone dimenzije eliptiqne nebule iznose 54\arcsec$\times$45\arcsec\ (pri \textrm{P.A.=69\fdg4}), {\ss}to je u veoma dobroj saglasnosti sa posmatra{\nj}ima ovog izvora u \textrm{H$\alpha$} i na 24\,\textrm{$\mu$m} u infracrvenoj oblasti. Koriste{\cc}i izmerene ugaone dimenzije i rastojanje od 3.94$\pm$0.27\,\textrm{kpc}, odredili smo da maglina ima linearne dimenzije od $1.03(\pm0.07)\times0.86(\pm0.06)$\,\textrm{pc}.

\newpage

{\ }

Na osnovu na{\ss}ih novih i ranije izmerenih vrednosti gustine fluksa, odredili smo spektralni indeks za Koenov Prsten od \textrm{$\alpha$=0.06$\pm$0.02}, {\ss}to ukazuje na to da je emisija Volf-Rajet magline termalnog porekla. Pretpostav{\lj}aju{\cc}i tipiqnu sred{\nj}u brzinu {\ss}ire{\nj}a od 75\,\textrm{km\,s$^{-1}$} za Volf-Rajet zvezdu sa crvenim supergigant progenitorom, procenili smo starost Koenovog Prstena na $\sim$13500~godina. Na{\ss}a posmatra{\nj}a \textrm{CO} emisije teleskopom NANTEN ukazuju na to da je Koenov Prsten nastao tokom relativno nedavnog doga{\dj}aja gubitka mase, dok pridru{\zz}eni molekulski gas koji emituje u \textrm{CO} linijama odra{\zz}ava kumulativne efekte povratnog dejstva zvezdanog vetra tokom znatno du{\zz}eg vremenskog perioda ($\sim$1.5\,\textrm{Myr}). Kao jedna od malobrojnih prstenastih Volf-Rajet maglina povezanih sa molekulskim gasom, Koenov Prsten predstavlja dragocenu laboratoriju za prouqavanje dugoroqnog uticaja povratnog dejstva masivnih zvezda na okolne molekulske oblake.

}
{\ }

%\end{onecolumn}

%\end{multicols}

%\end{strip}

%%%%%%%%%%%%%%%%%%%%%%%%%%%%%%%%%%%%%%%%%%%%%%%%%%%%%%%%%%%%%%%%%%%%%%%%%%%%%
%                       C O M P I L I N G
% You must compile your LaTeX source text "mysource.tex" in THE following
% way:
%  pdflatex mysource
%  bibtex mysource
%  pdflatex mysource
%  pdflatex mysource
% Compiling twice is required to set all the parameters needed
% (lastpage, labels...) properly. It will also generate the mysource.bbl
% with your bibliography.
%%%%%%%%%%%%%%%%%%%%%%%%%%%%%%%%%%%%%%%%%%%%%%%%%%%%%%%%%%%%%%%%%%%%%%%%%%%%%

\end{document}